# Controlling Laser Parameters by Electrical Polarization of Solid State Gain Media

Vladimir Chvykov

**Abstract:** The electrical polarization of the laser crystal by external electrical field can change significantly the output laser parameters such as wavelength of excitation and generation, spectral bandwidth, excitation and emission cross-sections, efficiency and so on. That can give new possibilities for wide control of these parameters, increase their range for existing lasers and create a new line of the laser devices. In this paper the direct impact of external electrical fields on the laser solid state gain media was estimated using as an example Ti:Sapphire crystal. It was demonstrated that metal-ligands distances displacement in this crystal under external electrical field of 20kV/cm can be comparable to that caused by mechanical pressure of 150 kbar, which was previously measured in experiments. The consequences of these displacements on the excitation and luminescence properties of the crystal are also discussed.

## Introduction

Controlling of the output laser parameters is a very important task for promoting new laser technology as well as laser devices applications. By now many efforts were entertained, most of which were concerning the optical parameters of oscillator's resonator or amplifier schemes or their optical elements. For example, implementing devices for controlling losses into laser cavity for reaching Q- switch or mod-lock regime [1-3], dispersive elements for tuning the output frequency, changing pump source parameters like spectral and temporal regime [4] and optical schemes and methods for pumping the laser amplifiers [5,6]. Nevertheless, the method proposed in this work for efficient control of the absorption and emission properties of the laser gain media themselves can help create a new line of laser devices due to higher efficiency, larger spectral bandwidth, tuning abilities and shorter pulse duration, higher repetition rates and pulse energy, quick switching between regimes of laser operation. Altogether, this can significantly increase the breadth of laser applications.

There are many works devoted to investigation of the absorption and emission property modification by implementation of mechanical pressure and temperature variation [7-10]. Nevertheless, these did not find large practical applications due to technological difficulties and long response time. From that point of view, strong electric and/or magnetic fields look more attractive. Well known and widely applicable for polarization rotation in crystals are Pockels and Kerr effects which change material polarization. Nevertheless, according our knowledge the attempts to research the direct impact of external electrical or magnetic fields on the laser solid state (SS) media were not entertained yet, excluding the semiconductor lasers.

There are a many SS laser media whose properties may be variable by strong electrical field, and particularly susceptible may be crystals doped with rare-earth and transition metal (TM) ions like $Ti^{3+}$, $Ni^{2+}$, $Cr^{2+}$, $Cr^{4+}$, $Mn^{5+}$. The former one possesses the optical electrons on the outer orbitals that will not be shielded from external electrical field by other ion orbitals as in the some rare-earth ions. The electron–vibration couplings between the electrons in ground and excited states, and host lattice are responsible for the emission Stokes shift and very broad band spectras. $Ti^{3+}$, among other TM, has the simplest electronic configuration ($3d^1$), outside of the closed-shell ions. Thus, as a first step, this work is devoted to research of the impact of strong electrical fields on radiation properties of Ti:Sapphire (Ti:Sa) crystals. Besides, this material is very promising for laser oscillators and master oscillator power amplifiers laser systems. Its high emission cross section ($4 \cdot 10^{-19}$ $cm^2$) and broad emission spectrum (~250 nm of FWHM) supports high single-pass gain, large tunability and 5 fs output pulses duration. The abilities successfully reduce losses even for very large crystal apertures and aspect ratios [11] resulted in a record output energy of ~200 J (~5 PW after compression) [12]. The high thermal conductivity of sapphire ~35W/(m·K) at room temperature increases by two orders of magnitude when cryogenically cooled, and is easily obtainable with refrigerators based on liquid nitrogen 77 K. Therefore, Ti:Sa systems are able to simultaneously achieve ultra-high peak as well as average power [13]. Moreover, the recent attempt of direct laser diode pumps [14] can further significantly increase the laser system efficiency.

**Ti:Sa crystal structure**

Ti:Sa crystal consists of the active $Ti^{3+}$ ions implemented as impurity in to sapphire ($Al_2O_3$) lattice. The impurity replace normally a few tenth of percent $Al^{3+}$ ions of the lattice weight. The elementary lattice cell composes octahedral structure where aluminum or titanium ions are surrounded by six oxygen $O^{2-}$ ions ($TiO_6$). The titanium ions possess a single "optical" electron on the outer shell which consists of five degenerated $d$-orbitals and 18 electrons on the filled-shells similar to a neutral argon atom. Location of the Ti-ion in the host lattice (octahedral cell) is resolving fivefold angular momentum degeneracy of $3d^1$-shell by local electrostatic field of neighboring $O^{2-}$ ions on the three ground $^2T_{2g}$ and two exited $^2E_g$ orbitals with energy difference $10Dq \sim 19130$ cm$^{-1}$, (where $10Dq$ is empirical parameter of crystal field electric strength) [10] lowering $^2T_{2g}$ by $-4Dq$ and raising $^2E_g$ by $+6Dq$ compared to initial level (Fig.1a). The location of electron on one of the three ground $T_{2g}$ orbitals further brakes the orbitals degeneracy on to $^2E_{2g}$ (consisting of two equivalent orbitals in $ZX$ and $ZY$ planes) and $^2B_{2g}$ (XY) due to changing the metal-ligand distances (Jahn-Teller (JT) effect) $\Delta_T^g \sim 670$ cm$^{-1}$ as well as exited $^2E_g$ orbitals on to $^2A_{1g}$ and $2B_{1g}$ with energy difference $\Delta_E^g \sim 2740$cm$^{-1}$. Optical excitation of the single electron on one of two exited $^2E_g$ orbitals - $^2A_{1g}$ or $^2B_{1g}$ orients along one or two directions on ligands does not change configuration of the former immediately due to their much heavier mass comparing to that of electron.

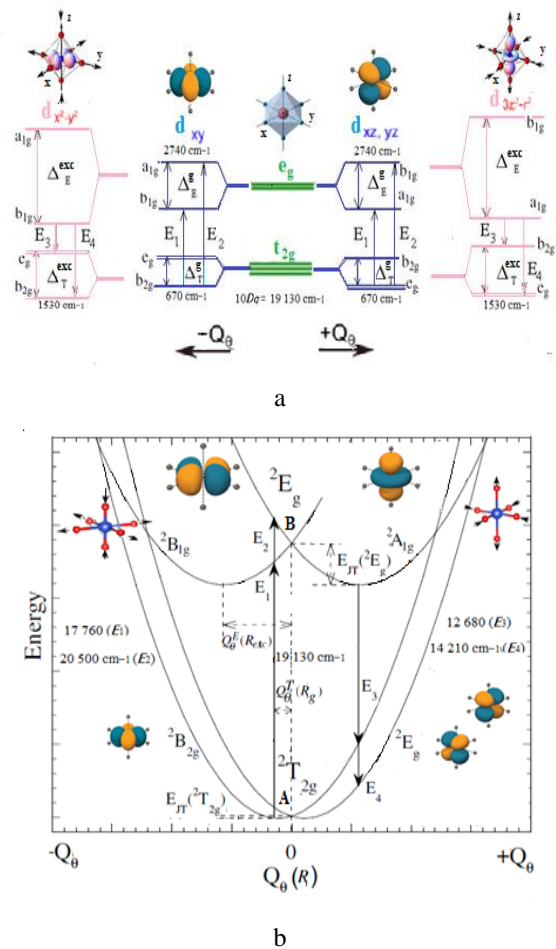

**Fig. 1 a-** Scheme of Ti$^{3+}$ ion energetic levels in sapphire host lattice; b - configuration-coordinate diagram of this ion in Ti:Sapphire crystal (modified picture from [10]).

Nevertheless, this configuration becomes unstable and appropriate *Ti-O* distances will be increased by ~0.2A⁰ [10] reaching new equilibrium positions and making octahedral cell compressed or elongated. This displacements further increase the energy difference between these threefold ground state orbitals $^2E_{2g}$ and $^2B_{2g}$, $\Delta_T^{exc}$ ~ 1530 cm$^{-1}$ and twofold exited ones $^2A_{1g}$ and $2B_{1g}$ with energy difference $\Delta_E^{exc}$ as well [Fig. 1a]. It also kicks the surrounding lattice and excites phonons with the energy coupling to the electronic energy levels of the Ti$^{3+}$ ions which is responsible for *Ti:Al$_2$O$_3$* absorption and luminescence Stokes-shift.

More detailed orbitals energy dependence on ions displacements can be presented on configuration-coordinate diagram (CCD), which is presented in Fig. 1 b. Schrodinger equation for adiabatic electron transmission described above can be obtained as [15]:

$$[H_e(r) + V_{el}(r)]\Psi_{n(R)}(r) = E_n(R) \Psi_{n(R)}(r) \qquad (1)$$

Here *r* and *R* represent electrons and ions positions consequently, $H_e(r)$ – Hamiltonian of electrons, $V_{el}(r)$ - represents the potential energy of electron and ion interactions, $\Psi_{n(R)}(r)$ –wave function of the n-th eigenstate with the energy $E_n(R)$. Since ions configuration doesn't change during transition, *R* remains a constants and can be varied as parameters. Adiabatic potential $U_n(R) = V_l(R) + E_n(R)$ of the n-state depending of these parameters varies continuously with *R* and can be presented as a function of atomic configuration on a CCD for the TiO$_6$ complex. Here $V_l(R)$ is potential energy of the lattice ions configuration. The shape of adiabatic potential curves for ground and exited states are different due to difference of geometrical equilibrium ligands positions as discussed above, so minimum of exited and ground potentials will be located on different configuration coordinates (average Ti-O distances) $R_{exc}$ and $R_g$. The relaxation of the exited state to this new position of equilibrium $R_{exc}$ is the origin of electron-lattice interaction. Its strength depends on the geometrical displacements $R_{exc}$ - $R_g$, the lattice relaxation energy, i.e. JT- stabilized energy $E_{JT}$ and derivative of adiabatic potential near equilibrium (generalized forces) ($\frac{dU_n}{dR}$). Lattice vibrations excited by this geometrical displacement $\Delta(R)$ in case of harmonic approximation can be transformed into normal modes: $\Delta R \sim \sum_k Q_k \exp(ikR)$ and then $U_n(\Delta R) = U_n(Q_k) \sim \sum_k \omega_k^2 Q_k^2$, *n* is ground or exited level, were *k* and $\omega_k$ correspondent wave vector and angular frequency of the normal vibration mode and $Q_k$ is normal coordinates.

Cross-section of CCD for the TiO$_6$ complex with respect of $Q_\epsilon = 0$, is presented on Fig. 1 b, where $Q_\theta$ and $Q_\epsilon$ is the two normal vibration mode coordinates which transform as $3z^2 - r^2$ and $x^2 - y^2$ respectively in octahedral symmetry. Threefold ground state orbitals $^2T_{2g}$ presents on the diagram two potential energy curves of $^2B_{2g}$ and $^2E_g$ symmetry orbitals with minimum stabilized energy $E_{JT}(^2T_{2g})$ are separated by $2Q_\theta^T(R_g)$ normal coordinate where $R_g$ as said above correspond to the average $Ti^{3+} - O^{2-}$ distances of the $TiO_6$ octahedron in the $^2T_{2g}$ ground state. Exited state orbitals $^2E_g$ are presented also by two curves $^2A_{1g}$ and $^2B_{1g}$ but with larger separation $2Q_\theta^E(R_{exc})$ and JT energy, where $R_{exc}$ is also the average $Ti^{3+} - O^{2-}$ distances but of the $^2E_g$ excited state. Large bandwidth excitation combines from two Gaussian peaks the corresponding two energy gaps $E_1$, $E_2$ (positions $Q_\theta^T(R_g)$ on the Fig.1 b). The optical transitions are represented as vertical lines in the figure because the configuration does not change during adiabatic excitation according Franck-Condon principle. After multiphonon relaxation to the minimal energy position of the one of two $^2E_g$ curves due to electron–vibration coupling, $Ti^{3+}$ ions is able generate very wide luminescence spectra also consisting of two large peaks with central wavelength equivalent $E_3$ and $E_4$ (positions $Q_\theta^E(R_{exc})$ on the Fig.1 b). Therefore, Ti:Sa crystal demonstrate the behaver of 4-level laser system due to much faster vibrational relaxation rate about few hundreds of fs, compared to the exited level lifetime ~ 3.2μs.

## Modification of the crystal geometry

It is clear from above explanation that one can modify the absorption and/or emission parameters of Ti:Sa laser crystal reliably by changing the distances between *Ti*- ion and ligands. For example, the uniform reduction of these distances [10] will increase the cubic field component of the local electrical field of crystal (*Dq*), JT- energy and make a stronger electron-phonon coupling. On CCD the energy difference at 0-equilibrium position will be increased (distance between A and B points), as well as $E_{JT}(^2T_{2g})$ and $E_{JT}(^2E_g)$ and zero-phonon positions of $^2T_{2g}$ and $^2E_g$ curves ($Q_\theta^T(R_g)$ and $Q_\theta^E(R_{exc})$). Consequently, this will lead to the blue shift of the excitation spectra, red-shift of the emission spectra and bandwidth enlargement of both. One can expect the opposite results for the case of uniform increase of

Ti-O distances. More complicated consequences can be anticipated if asymmetric distortion of the $TiO_6$ cell will be possible [9]. That will bring also the asymmetrical distortion of the CCD curves branches of both ground and exited levels including, besides others effects, the changing of generalized forces ($\frac{dU_n}{dR}$) and so graph curvatures, and as result, possible large scale modifications of excitation-emission parameters.

Most direct way to modify the metal-ligands distances is the application of mechanical pressure on crystal. Multiple publications were devoted to these researches. The investigations of uniform hydrostatic as well as uniaxial (parallel C-axis) compressive stress influence on luminescence and absorption properties of Ti:Sa crystals was presented in [7-10]. In case of the hydrostatic pressure the excitation blue-shifts were demonstrated up to 1000 cm$^{-1}$, and Dq-parameter increasing up to 120cm$^{-1}$ when the pressure was variated from 0 to 150 kbar [7], as well as the excitation ~ 680cm$^{-1}$ and emission spectra ~470cm$^{-1}$ blue-shifts and the increasing lifetime of the exited levels on 0.6 μs when pressure was increased from atmospheric to 80 kbar [10]. At the same time, the uniaxial compressive stress leads to blue–shift (70-90 cm$^{-1}$) of the excitation spectra and red-shift (below 10 cm$^{-1}$) of the emission one when uniaxial compressive stress was changed from 0 to 20 kbar [9].

It is important for further evaluation to estimate the possible shift of the Ti-O distances under such pressure. That can be done using the measurements corresponding Dq parameter dependence on pressure [7] and taking into account that this value should vary as $R^{-5}$ for the approximation of a point charge ligand [16] and as $R^{-6}$ for point dipoles, [17] where R is the distance from the transition metal ions to the ligands. It was shown, Dq changing was equal 120 cm$^{-1}$ after undergoing 140 kbar of pressure. On the other hand, $10Dq = 19130$ cm$^{-1}$ and the average $Ti – O$ distance on the ground level is 1.9A$^0$ [10]. From the relation $R_p/R = (Dq_p/Dq)^{1/6}$, where $R_p$ Ti – O distance and $Dq_p$ parameter means under the pressure, one can realize $\Delta R$ =0.03 A$^0$ which is about 5 times smaller than this distance variation during excitation (0.16 A$^0$). The authors claimed the abilities to modify the laser parameters using these techniques, but until now, this did not find wide application in practice because of the difficulties to embed the stress devices into the laser oscillators and amplifiers, very limited crystal sizes and long response times.

**Displacement due to external electric field.**

The influence of external electric field onto the laser crystals could give more promising approaches compared to the mechanical pressure discussed above. The high voltage could be applied and switched off during a very short (sub-ns) time with the quick reconnection if need, and a very precise timing to pump light (few ps), in any crystal directions and with very high amplitude (as close as desired to breakdown of the crystal, which is for Ti:Sa, for example, is about 480 kV/cm). The technology of the crystals polarization by the high voltage is well developed, for instance, for Pockels cells. On the other hand, the electrical polarization of the crystal will lead to the asymmetrical distortion and as it was discussed above, may bring more benefits for the modification of the emission/excitation parameters.

Now it is worth to estimate the possible Ti-O length distortion due to reasonable external electrical field and to compare it with this distance shifted by the light excitation and mechanical pressure to be sure that the electrical distortion will be enough to make a sufficient influence on the crystal lasing parameters. This estimation can be done roughly using the well-known Classius-Mosotti relation between the macro and micro polarization parameters in dielectrics. $\frac{Na}{3\epsilon_0} = \frac{\epsilon_r-1}{\epsilon_r+2}$ , where N is concentration of the ions in sapphire, $\epsilon_0$ – permittivity free space, $\epsilon_r$ – dielectric constant, and $a$ – polarizability constant. Taking $N$ =0.23 x 10$^{23}$ cm$^{-3}$ and $\epsilon_r$ = 10 for sapphire one can get $a$ = 0.86 x 10$^{-31}$ Fcm$^2$. On the other hand, $aE = \mu = \Delta xq$, where E is the strength of electrical field, $\mu$- dipole moment, $q$ – charge, which is $3e$ for TiO$_6$ ($e$-charge of electron) and $\Delta x$ - charge displacement. Polarizability constant ($a$) for these crystals consist of the electronic ($a_e$) and ionic ($a_i$) constants, nevertheless, $a_e$ is negligible (~ 10$^{-36}$ Fcm$^2$) and can be omitted in this estimation. For modeling the cell $TiO_6$ with the $Ti^{3+}$ ions charge $+3e$ one can take the equivalent dipole, taking into account equivalent charge from each $O^{2-}$ ions consisting the cell meaning the rest of their charges involved in other cells. Thus, one can derivate the dependence of $\Delta x$ on E: $\Delta x = \frac{\epsilon_0}{Ne} \frac{\epsilon_r-1}{\epsilon_r+2} E$ or $\Delta x$= 0.18•10$^{-4}$ E [A$^0$] after calculation. Thus, if the external electrical field of 20kV/cm will be applied to crystal the dipole displacement $\Delta x$ = 0.36 A$^0$ can be expected, The equivalent average Ti-O displacements could be estimated due to their projection on to the electric field direction Fig. 2 a. and is $\Delta R = 0.034$ A⁰ which is close to the ligand displacement under 150 kbar mechanical pressure (see calculation above).

The consequences of this deformation can be evaluated by the effect of the corresponding CCD modification. Using the harmonic approximation discussed above, accepting the terminology of CCD on the Fig. 2 b and making few simplifications $\{Q_\theta^E(R_{exc}) = Q_\theta^E(R_{exc}) - Q_\theta^T(R_g)$ and $Q_\theta^T(R_g) = 0$ due to $Q_\theta^T(R_g) << Q_\theta^E(R_{exc})\}$ and zero phonon transition $E(^2T_{2g})=0$ due to $E(^2T_{2g})<<10Dq)$ one can write $E_g(Q_\theta)=a_g Q_\theta^2$ and $E_{exc}(Q_\theta)=a_{exc}(Q_\theta - Q_\theta^E(R_{exc}))^2 + 10Dq$ were $a_g \sim \omega_\theta^2$ and $a_{exc} = ka_g$, $k$ is arbitrary parameter. Subtracting $E_g(Q_\theta)$ from $E_{exc}(Q_\theta)$, taking $Q_\theta=0$ and $Q_\theta = Q_\theta^E(R_{exc})$ one can get excitation and emission energy:

$$E_{exc}=10Dq + 4ka_g(Q_\theta^E(R_{exc}))^2$$
$$E_{em}=10Dq - 4a_g(Q_\theta^E(R_{exc}))^2 \qquad (2)$$

The spectral bandwidth could be found taking $\Delta Q_\theta = 0\pm\delta(Q_\theta)$ for excitation and $\Delta Q_\theta = Q_\theta^E(R_{exc})\pm\delta(Q_\theta)$, for emission:

$$\Delta E_{exc} = 4ka_g Q_\theta^E(R_{exc})\delta(Q_\theta)$$
$$\Delta E_{em} = 4a_g Q_\theta^E(R_{exc})\delta(Q_\theta). \qquad (3)$$

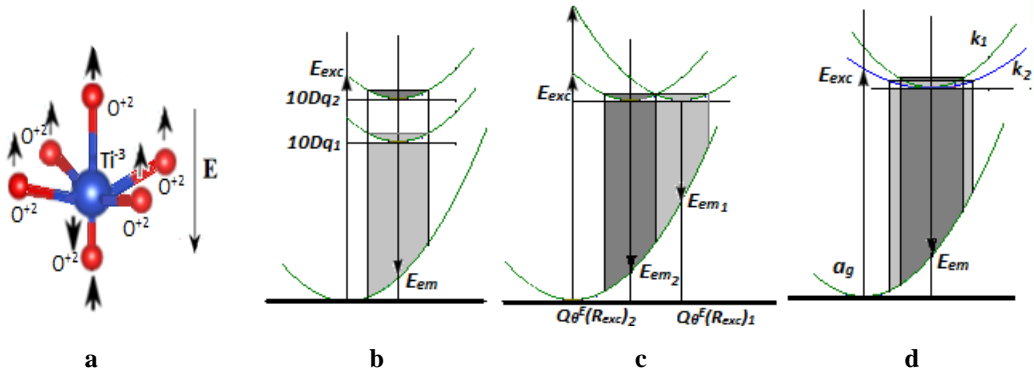

**Fig. 2 a-** model of TiO₆ cell under the electric field influence, **b-** CCD distortion of the Ti- ion electronic states under two different local cubic electrical field, **c –** the same for two different zero phonon positions, **d-** the same for two different strengths of photon-phonon coupling.

The changing of these four parameters, namely, $10Dq$, $Q_\theta^E(R_{exc})$, $a_g$ and $k$ one can get the reliable modifications in the excitation and emission spectras. It could be worth to consider the separate influence of each of these parameters using the formulas (2 and 3) for $E_{exc}$, $E_{em}$, $\Delta E_{exc}$ and $\Delta E_{em}$ derived above. First case is the variation of $10Dq$ (Fig. 2 b), keeping for simplicity all other parameters constants. As seen from formulas and figure, $E_{exc}$ and $E_{em}$ are increasing or reducing together with $10Dq$ and $\Delta E_{exc}$ and $\Delta E_{em}$ doesn't change. Second case (Fig. 2 c) is the $Q_\theta^E(R_{exc})$ variation: if this parameter is increasing, the $E_{exc}$ is increasing too (blue-shift of the central wavelength) and $E_{em}$ reducing (red-shift) and $\Delta E_{exc}$, $\Delta E_{em}$ both experience the same rate of growth. The final case is the variation of $k$-parameter, which changes the curvature of the exited state level comparing to the ground state. The central excitation energy will drop down with reduction of this parameter, when emission energy doesn't depend on $k$ and so, doesn't change. For $\Delta E_{exc}$ and $\Delta E_{em}$ analysis one should take in to account the dependence $\delta(Q_\theta)$ on $k$-parameter with the assumption of population inversion distribution on the upper level remaining constant during this changing curvature. In the case of Boltzmann distribution $\delta(Q_\theta) = (k_b T/ ka_g)^{1/2}$ were $k_b$ - Boltzmann constant and $T$-temperature, then

$$\Delta E_{exc} = 4(k_b T k a_g)^{1/2} Q_\theta^E(R_{exc})$$
$$\Delta E_{em} = 4 (k_b T k a_g)^{-1/2} Q_\theta^E(R_{exc}), \qquad (4)$$

Which means $\Delta E_{exc} \sim (k)^{1/2}$, reducing excitation spectra bandwidth and $\Delta E_{em} \sim (k)^{-1/2}$, increasing of the emission one with the reduction of $k$-parameter.

Keeping in a mind the possible laser parameter modification by variation of the gain media described above, one can expect the efficiency improvement due to the quantum defect variation ($E_{exc}/E_{em}$), reducing $E_{exc}$ and allowing to use more efficient pump lasers (semiconductor lasers, for example), increasing emission bandwidth ($\Delta E_{em}$) and reducing the pulse duration and variation of the excitation/emission cross section. As seen from presented discussion, the quantum defect, for example, could be reduced due to the decreasing $2Q_\theta^E(R_{exc})$ separation, which can be done by elongation of the corresponded Ti-O distances and/or by reducing *k*-factor. The red-shift of the excitation is possible with reduction of the *k*-factor also or 10Dq-parameter (cubic local field in $TiO_6$ cell), which can be decreased by elongation of the average Ti-O distance. The emission bandwidth could be increased by decreasing *k*-factor, and also by elongation of the average Ti-O distance, which consequently leads to reduced photon-phonon coupling. The variation of excitation/emission bandwidth also will impact the variation excitation/emission cross section and so on. The detailed numerical simulations of these effects can be done after delivering all the required experimental data.